\documentclass[italian,english]{article}
\usepackage[latin9]{inputenc}
\usepackage{color}
\usepackage{graphicx}
\usepackage{amssymb}
\usepackage{esint}

\makeatletter
\newcommand{\lyxaddress}[1]{
\par {\raggedright #1
\vspace{1.4em}
\noindent\par}
}

\makeatother

\usepackage{babel}

\begin{document}

\title{\textbf{Radiation dominated era} \textbf{and the power of general
relativity}}

\author{\textbf{Christian Corda}}

\maketitle
\begin{center}
Institute for Theoretical Physics and Advanced Mathematics Einstein-Galilei,
Via Santa Gonda 14, 59100 Prato, Italy 
\par\end{center}

\begin{center}
and 
\par\end{center}

\begin{center}
Inter-University Centre Engineering of life and Environment, LIUM
University, Via Lugano 2, 6500 Bellinzona, Switzerland
\par\end{center}

\lyxaddress{\begin{center}
\textit{E-mail addresses:} \textcolor{blue}{cordac.galilei@gmail.com} 
\par\end{center}}
\begin{abstract}
An analysis in the framework of the radiation dominated era permits
to put bounds on the weak modification of general relativity which
arises from the Lagrangian $R^{1+\varepsilon}$. Such a theory has
been recently discussed in various papers in the literature. The new
bounds together with previous ones in the literature rule out this
theory in an ultimate way.
\end{abstract}
It is well known that Einstein's General Relativity Theory (GRT) can
be adopted to describe various astrophysical observations and, at
scales of the Solar System, it results consistent with many, very
accurately precise, astrophysical measurements, such as the gravitational
bending of light, the perihelion precession of Mercury and the Shapiro
time delay \cite{key-1,key-2}. On the other hand, at larger scales,
several shortcomings are present, like the famous Dark Energy \cite{key-3}
and Dark Matter \cite{key-4} problems.

An alternative approach consists in assuming that gravitational interaction
could act in different way at large scales \cite{key-5}. This different
framework does not require to find out candidates for dark energy
and dark matter at fundamental level (not detected up to now), but
takes into account only the observed ingredients (i.e. curvature,
radiation and baryon matter), changing the left hand side of the field
equations \cite{key-6,key-7}. In this way, a room for alternative
theories can be introduced and the most popular Dark Energy and Dark
Matter models can be, in principle, achieved by considering Extended
Theories of Gravity (ETG), i.e $f(R)$ theories of gravity, where
$R$ is the Ricci curvature scalar, see \cite{key-6}-\cite{key-10},
 \cite{key-31,key-32} and references within, and Scalar Tensor Teories
\cite{key-11,key-12,key-13}, which are generalizations of the Jordan-Fierz-Brans-Dicke
Theory \cite{key-14,key-15,key-16}. 

An ultimate endorsement for the approach of ETG should be the realization
of a consistent gravitational wave astronomy \cite{key-11}. In fact,
in the case of ETG, gravitational waves generate different oscillations
of test masses with respect to gravitational waves in standard GRT.
Thus, an accurate analysis of such a motion can be used in order to
discriminate among various theories, see \cite{key-11} for details.

Another key point is that Solar System tests imply that modifications
of GRT in the sense of ETG have to be very weak \cite{key-5,key-11}.
In other words, such theories have to be \emph{viable}. In the framework
of viable ETG, the theory arising from the action (in this paper we
work with $8\pi G=1$, $c=1$ and $\hslash=1$)

\begin{equation}
S=\frac{1}{2}\int d^{4}x\left(\sqrt{-g}f_{0}R^{1+\varepsilon}+S_{m}\right),\label{eq: high order 1}\end{equation}

where $f_{0}>0$ has the dimensions of a mass squared and $\varepsilon$
is a small real number, has been discussed in various papers in the
literature \cite{key-17}-\cite{key-27} and \cite{key-33}. Equation
(\ref{eq: high order 1}) is a particular choice in $f(R)$ theories
of gravity \cite{key-6}-\cite{key-10},  \cite{key-31,key-32} with
respect to the well known canonical one of General Relativity (the
Einstein - Hilbert action \cite{key-28}) which is 

\begin{equation}
S=\frac{1}{2}\int d^{4}x\left(\sqrt{-g}R+S_{m}\right).\label{eq: EH}\end{equation}

Various observational constraints set the limits \begin{equation}
0\leq\varepsilon\leq7.2\times10^{-19}\label{eq: limits}\end{equation}
on the parameter $\varepsilon$ \cite{key-17,key-18}, while the recent
work \cite{key-25} obtained a lower limit

\begin{equation}
0\leq\varepsilon\leq5\times10^{-30}.\label{eq: limits-1}\end{equation}

Gravitational waves in this particular theory have been discussed
in \cite{key-26}. In \cite{key-27}, a spherically symmetric and
stationary universe has been analyzed in the tapestry of this theory.

In order to discuss this particular theory in the framework of the
radiation dominated era, the well known Friedman-Robertson-Walker
cosmological line - element has to be used \cite{key-1,key-28}, and,
for the sake of simplicity, we will consider the flat case, because
the WMAP data are in agreement with it \cite{key-29}\begin{equation}
ds^{2}=-dt^{2}+a^{2}(dz^{2}+dx^{2}+dy^{2}).\label{eq: metrica FRW}\end{equation}

We also recall that in the radiation dominated era the equation of
state is \cite{key-1}

\begin{equation}
p=\frac{1}{3}\rho\label{eq: equation of state}\end{equation}

and the the energy density is given by \cite{key-1}

\begin{equation}
\rho=\frac{f\pi^{2}}{120}k^{4}T^{4},\label{eq: energy density}\end{equation}

where $k$ is the Boltzmann constant and $f$ is a parameter depending
from the particular radiation, for example $f=8$ for electromagnetic
radiation, $f=7$ for neutrinos, etc., see \cite{key-1} for details.

By varying the action (\ref{eq: high order 1}) with respect to $g_{\mu\nu}$
(see \cite{key-26} for a detailed computation) the field equations
are obtained \begin{equation}
G_{\mu\nu}=\frac{1}{(1+\varepsilon)f_{0}R^{\varepsilon}}\{-\frac{1}{2}g_{\mu\nu}\varepsilon f_{0}R^{1+\varepsilon}+[(1+\varepsilon)f_{0}R^{\varepsilon}]_{;\mu;\nu}-g_{\mu\nu}\square[(1+\varepsilon)f_{0}R^{\varepsilon}]\}+T_{\mu\nu},\label{eq: einstein 2}\end{equation}

where \begin{equation}
T_{\mu\nu}=\left|\begin{array}{cccc}
\rho & 0 & 0 & 0\\
0 & p & 0 & 0\\
0 & 0 & p & 0\\
0 & 0 & 0 & p\end{array}\right|\label{eq: stress-energy}\end{equation}

is the well known stress-energy tensor of the matter \cite{key-1,key-28}.
Taking the trace of the field equations (\ref{eq: einstein 2}) one
gets 

\begin{equation}
3\square(1+\varepsilon)f_{0}R^{\varepsilon}=(1-\varepsilon)f_{0}R^{1+\varepsilon}+T,\label{eq: KG}\end{equation}

where $T=\rho-3p$ is the trace of the stress-energy tensor (\ref{eq: stress-energy})
\cite{key-28}.

Following \cite{key-1}, if one computes the components of eqs. (\ref{eq: einstein 2})
and (\ref{eq: KG}) by using the line element (\ref{eq: metrica FRW})
three independent Friedman equations are obtained

\begin{equation}
\begin{array}{c}
3\dot{R}^{2}(1-\varepsilon^{2})\varepsilon f_{0}R^{\varepsilon-2}+3\ddot{R}(1+\varepsilon)\varepsilon f_{0}R^{\varepsilon-1}+9\frac{\dot{a}}{a}\dot{R}(1+\varepsilon)\varepsilon f_{0}R^{\varepsilon-1}-2f_{0}R^{1+\varepsilon}=0\\
\\6\left(\frac{\ddot{a}}{a}+\frac{\dot{a}^{2}}{a^{2}}\right)(1+\varepsilon)f_{0}R^{\varepsilon}+3f_{0}R^{1+\varepsilon}-12\frac{\dot{a}}{a}\dot{R}(1+\varepsilon)\varepsilon f_{0}R^{\varepsilon-1}+6\dot{R}^{2}(1-\varepsilon^{2})\varepsilon f_{0}R^{\varepsilon-2}\\
\\6\frac{\ddot{a}}{a}(1+\varepsilon)f_{0}R^{\varepsilon}+f_{0}R^{1+\varepsilon}-6\frac{\dot{a}}{a}\dot{R}(1+\varepsilon)\varepsilon f_{0}R^{\varepsilon-1}=\rho.\end{array}-6\ddot{R}(1+\varepsilon)\varepsilon f_{0}R^{\varepsilon-1}=\rho\label{eq: Friedman}\end{equation}

We recall that the Ricci scalar is given by \cite{key-30} \begin{equation}
R=-6[\frac{\ddot{a}}{a}+(\frac{\dot{a}}{a})^{2}].\label{eq: Ricci Scalar}\end{equation}
One can also use the Bianchi identities \cite{key-1} to get another
independent equation

\begin{equation}
a\dot{\rho}=-4\rho\dot{a}.\label{eq: bianchi}\end{equation}

In standard general relativity, during the radiation dominated era,
the scale factor is \cite{key-1}

\begin{equation}
a\sim t^{\frac{1}{2}}.\label{eq: scala}\end{equation}

Hence, in the theory which arises from the action (\ref{eq: high order 1})
one assumes

\begin{equation}
a\sim t^{(\frac{1}{2}+\delta)}.\label{eq: scala 2}\end{equation}

By using the second and the third of eqs. (\ref{eq: Friedman}) and
eq. (\ref{eq: bianchi}) one gets

\begin{equation}
\varepsilon=2\delta.\label{eq: epslon}\end{equation}

By deriving eq. (\ref{eq: Ricci Scalar}) and by using eq. (\ref{eq: epslon})
we write

\begin{equation}
\dot{R}=\frac{6\varepsilon(1+\varepsilon)}{t^{3}}.\label{eq: R punto}\end{equation}

Considering the third of eqs. (\ref{eq: Friedman}) together with
eq. (\ref{eq: energy density}) one obtains

\begin{equation}
T=\sqrt[4]{\frac{60}{4\pi^{3}f}}\sqrt[4]{6^{(1+\varepsilon)}(\frac{1+\varepsilon}{2}){}^{(1+\varepsilon)}\frac{1}{4}\left[8(1+\varepsilon)-5(1+\varepsilon)^{2}-2\right](-\varepsilon)^{\varepsilon}}\frac{1}{kt^{\frac{(1+\varepsilon)}{2}}}.\label{eq: T}\end{equation}

Putting

\begin{equation}
F(\varepsilon)\equiv6^{(1+\varepsilon)}(\frac{1+\varepsilon}{2}){}^{(1+\varepsilon)}\frac{1}{4}\left[8(1+\varepsilon)-5(1+\varepsilon)^{2}-2\right](-\varepsilon)^{\varepsilon},\label{eq: F di epslon}\end{equation}

we need \begin{equation}
F(\varepsilon)\geq0\label{eq: reale}\end{equation}

in order $T$ to be a real value. The constrain (\ref{eq: reale})
is satisfied for 

\begin{equation}
-0.69\leq\varepsilon\leq0,\label{eq: final bound}\end{equation}

see figure 1. 

\begin{figure}
\caption{The function $F(\varepsilon)$}

\includegraphics[scale=0.7]{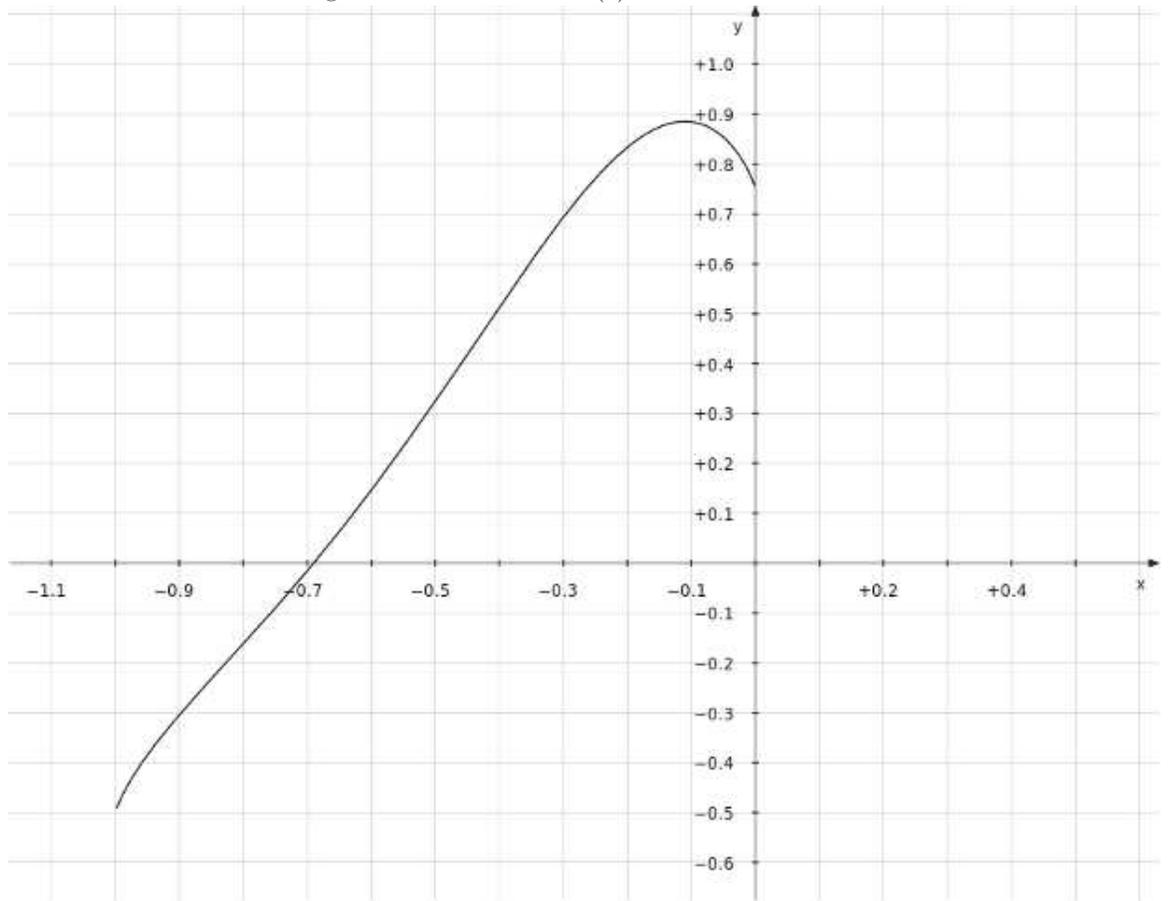}

The function $F$ is plotted. The x axis represents the variable $\varepsilon,$
the y axis the variable $F(\varepsilon)$. We see that the condition
$F(\varepsilon)\geq0$ is satisfied for $-0.69\leq\varepsilon\leq0.$
\end{figure}

Considering the bound (\ref{eq: final bound}) togheter with the bounds
(\ref{eq: limits}) and (\ref{eq: limits-1}) one gets immediately
$\varepsilon=0,$ i.e. general relativity is recovered and the theory
which arises from the action (\ref{eq: high order 1}) is ultimately
ruled out.

In summary, in this work we realized an analysis in the framework
of the radiation dominated era in order to put bounds on the weak
modification of general relativity which arises from the action (\ref{eq: high order 1}).
The new bounds together with previous ones in the literature rule
out this theory in a definitive way.

\subsubsection*{Acknowledgements}

I thank a reviewer for useful comments.
\selectlanguage{italian}%

\end{document}